# Spin-Dependent Charge-Carrier Recombination Processes in Tris(8-Hydroxyquinolinato) Aluminum

H. Popli,[1] X. Liu,[1] T. H. Tennahewa,[1] M. Y. Teferi,[1] Evan Lafalce,[1] H. Malissa,[1] Z. V. Vardeny,[1] and C. Boehme[1]

[1]*Department of Physics and Astronomy, University of Utah, Salt Lake City, Utah 84112, USA*

## Abstract

We have studied the nature and dynamics of spin-dependent charge carrier recombination in Tris(8-hydroxyquinolinato) aluminum ($Alq_3$) films in light emitting diodes at room temperature using continuous wave and pulsed electrically detected magnetic resonance (EDMR) spectroscopy. We found that the EDMR signal is dominated by an electron-hole recombination process, and another, weaker EDMR signal whose fundamental nature was investigated. From the pulsed EDMR measurements we obtained a carrier spin relaxation time, $T_2 = 45\pm25$ ns which is much shorter than $T_2$ in conjugated polymers, but relatively long for a molecule containing elements with high atomic number. Using multi-frequency continuous wave EDMR spectroscopy, we obtained the local hyperfine field distributions for electrons and holes, as well as their respective spin-orbit coupling induced *g*-factor and *g*-strain values.

I. INTRODUCTION

The interest in organic semiconductor (OSEC)-based spintronics devices [1–3] has led to extensive studies of these materials' opto-electronic [4] and magneto-electronic properties [5,6]. Electronic excitations in OSECs generally exhibit weak spin-orbit coupling (SOC), and, consequently, charge carrier transport, recombination, and spin transport mechanisms can be dominated by spin-selection rules. There are ample examples of such spin-dependent processes [7–14] in OSECs, which involve localized spin-1/2 carriers or polarons ($P^+$ and $P^-$) [7], weakly coupled polaron pairs ($P^+P^-$) [11,15], as well as more complex paramagnetic electronic excitations such as trions and spin triplet-polaron complexes [16]. One important experimental technique for the study of such spin-dependent processes in OSECs have been electron paramagnetic resonance-based techniques including optically (ODMR) and electrically (EDMR) detected magnetic resonance spectroscopies [9,16–18].

Here we present a study of spin-dependent electronic transitions in thin solid films based on tris-(8-hydroxyquinolinato) aluminum ($Alq_3$), a small molecule that consists of a central aluminum atom with three side groups ($C_{27}H_{18}AlN_3O_3$). $Alq_3$ has been used as an electroluminescent layer in high-efficiency organic light emitting diodes (OLEDs) [19] and other applications such as spin valves [20] and light emitting transistors [21]. In the past, the nature of charge carriers and spin-dependent charge carrier recombination in $Alq_3$-based OLEDs has already been studied with continuous wave (cw) EDMR and ODMR spectroscopies [22–24], revealing that charge carriers in $Alq_3$ films form polarons, and the precursor excitations to excitons generation are polaron-pairs [24]. It has also been claimed that other excitations such as bipolarons ($BP^{++}$ and $BP^{--}$) and triplet excitons may contribute to spin-dependent recombination [25,26].

Following our previous studies of spin-dependent electronic processes in OLEDs based on π-conjugated polymers using transient ODMR techniques [8,27,28], the study presented here has been motivated by the prospect to apply a broad regimes of multi-frequency and pulsed, coherent spin manipulation techniques to the study of $Alq_3$ based devices. As $Alq_3$ is a small molecule and not conjugated polymer, and it involves a high density of an elements with high atomic number (Al), we hypothesize that electronic spin states in $Alq_3$ exhibit much higher SOC, and this may affect the physical behavior of spin-dependent processes here, in contrast to those in π-conjugated polymers. Using pulsed and multi-frequency EDMR spectroscopies [29], including electrically detected Hahn-echo experiments, electrically detected spin-Rabi oscillation measurements and other coherent spin-motion experiments [11,30], we have studied charge carrier spin relaxation times and compare them to other OSECs with high atomic numbers, e.g. Pt-rich polymers [31].

## II.    EXPERIMENTS

OLEDs with active layers of Alq$_3$ were prepared on glass templates with lithographically defined thin-film ITO wiring as described by McCamey et al. [29]. For their preparation, a 50 nm thick layers of PEDOT:PSS, which serve as a hole injectors, were deposited on ITO contact pads by spin-casting and these subsequently annealed at 110ºC for 15 min. Alq$_3$ layers with thicknesses of ~30 nm were then deposited by thermal evaporation for about 3 mins at a base pressure of $\sim 10^{-7}$ mbar using a tungsten basket and quartz cuvette in the evaporator. Finally, 7 nm thick calcium electron injection layers and 100 nm aluminium top electrodes were thermally evaporated, before these devices were encapsulated with SiO$_2$ insulating layers. The finished devices emitted green electroluminescence under application of sufficient bias voltages. In order to ensure bipolar charge carrier injection, all devices were characterized by measurements of their I-V characteristics using a Keithley 2400 source-meter. In Figure 1(a) the used Alq$_3$ OLED layer structure, a picture of one OLED device under operation, the luminescence spectrum of Alq$_3$, as well as a representative I-V curve are shown.

First, EDMR experiments were carried out using a commercial Bruker ElexSys 580 spectrometer. For the pulsed operation, short high-power microwave pulses were applied to the device under operation in the FlexLine ER 4118X-MD-5W1 dielectric resonator inside an Oxford CF935 cryostat in order to allow for measurements between 5 K and room temperature. The device bias voltage was provided by a Stanford Research Systems SIM928 isolated battery source, that was adjusted to voltages ranging from 3 to 20 V, while the OLED current was measured by a SR570 transimpedance amplifier with a gain of 2 µA/V. The experimental setup is schematically depicted in Fig. 1(b). For measurements of current transients following the MW excitation as a function of magnetic field [8,16,29,32], the output voltage of the current amplifier was recorded by the built-in fast digitizer of the E580 system. In this setup, a bandpass filter with cut-off frequency at 30 Hz and 30 kHz is adjusted within the transimpedance amplifier. The measurement sequence was repeated at a shot-repetition rate of 0.454 kHz.

In order to observe the characteristics of coherent charge-carrier spin motion in pulsed EDMR measurements, we first conducted electrically detected charge-carrier spin-Rabi oscillation measurements. In such a measurement scheme, the device current—measured using the SR570 current amplifier with a 30 Hz and 30 kHz bandpass filter—is typically digitally integrated by the E580 transient recorder over an interval suitable for the dynamics of the studied spin-dependent processes, i.e. for the studied Alq3 processes reported here for 16.38 µs, starting 2 µs after the MW pulse in order to account for the rise time of the detector [11,29,30,32,33]. For increased signal-to-noise ratio, the device current can also be integrated through an analog boxcar integrator (Stanford Research Systems SR250) that is triggered and read-out by the E580 system. The integration interval is 15 µs long and starts 2 µs after the read-out pulse. In this case, the transimpedance amplifier uses a high-pass filter with a cut-off frequency of 10 Hz and a gain of 20 µA/V. The duration of the MW pulse is incremented at each step to achieve increasing flip angles. This measurement sequence is depicted in the inset of Fig. 3(a).

For measurements of the spin coherence times of charge-carrier pairs in Alq$_3$ OLEDs, we conducted electrically detected Hahn echo measurements, where we applied a modified $\pi/2 - \tau' - \pi - \tau - \pi/2 - \int Idt$ pulse sequence that is described in detail in Refs. [8,33–37]. The final $\pi/2$ readout-pulse within this sequence is required in order to project the spin ensemble onto their eigenstates along the direction of the external magnetic field, where the echo amplitude is detected as charge signal. The echo shape is detected by varying $\tau$ with respect to $\tau'$, with the echo maximum occurring at $\tau = \tau'$. The echo sequence is illustrated in Fig. 3(e).

Multi-frequency continuous wave EDMR measurements were conducted using a series of EDMR measurement robe-heads which fit inside the Oxford CF935 cryostat, and which employ coplanar waveguide resonator structures and NMR-style RF coils for electron paramagnetic resonant excitation over a wide range of excitation frequencies. Multi-frequency EDMR experiments allow to distinguish frequency-dependent and frequency-independent line broadening effects [8,27,28,34,38,39]. These experiments were also carried out in the E580 spectrometer, using a separate MW source (Agilent N5181A for frequencies below 6 GHz and Agilent N5173B for frequencies above 6 GHz), as well as a MW amplifier (Mini-Circuits ZHL-5W-1) that was directly connected to the probe-head. For these multi frequency EDMR experiments, changes in device currents were detected, again using the SR570 transimpedance amplifier, with a sensitivity of 2 µA/V and a band pass filters with upper and lower cut-off frequencies of 30 Hz and 30 kHz, respectively. For these multi-frequency experiments, the output of the current amplifier was connected to the input of the lock-in amplifier that is part of the E580 system. We used amplitude modulated lock-in detection [17] with source frequencies between 100 MHz and 5.55 GHz, and magnetic field modulated lock-in detection for source frequencies between X-Band and 16.885 GHz.

### III.    RESULTS

Figure 2(a) displays a color plot of the room temperature device current changes from the steady state, following a 400 ns long MW pulse at a MW power of 1 kW as a function of time after the excitation pulse at $t = 0$ (horizontal axis) and the magnetic field (vertical axis) in a magnetic field range that includes, for the given frequency, the magnetic resonance of vacuum electrons ($g$=2.0023..) at ~343.6 mT. The steady-state device current for these measurements was 50 µA. In the displayed data set, non-resonant, i.e.

magnetic-field independent current changes due to the pulsed excitation, have been subtracted through a zeroth order baseline correction, i.e. a constant. The data shows a pronounced magnetic resonant response at $g \sim 2$, consisting of a signal that displays a short initial current enhancement, followed by a long, slow quenching which lasts well beyond the recorded time range of 100 µs.

Using this signal, we conducted coherent transient nutation measurements, [11,29,30] at room temperature [cf. Fig. 3(a)] and at 5 K [cf. Fig. 3(c)]. The duration of the strong coherent excitation (1kW nominal microwave power) was varied between 0 ns and 60 ns in steps of 2 ns. The time-integrated current responses, i.e. the charges transmitted due to the pulsed excitations, were then recorded as functions of the excitation length. The measurement sequence is shown in the inset of Fig. 3(a). These experiments reveal an oscillatory behavior of the integrated spin-dependent currents, which, however, decay rather fast: at room temperature [cf. Fig. 3(a)] the oscillations have decayed to the noise level within at ~3 cycles, while, at low temperatures (5 K) [cf. Fig. 3(b)] ~5 cycles are discernible. The envelope of the oscillations fades with a time constant of $T_2^*$, the dephasing time, which is found to be 20 ns at room temperature, and 55 ns at 5 K.

Since $T_2^*$ places a lower limit on the spin-spin relaxation time $T_2$, we attempted to measure electrically detected Hahn echoes [8,33–37]. The measurement sequence is shown in Fig. 3(e). The measurements showed that echo signals were not discernible for pulse delays down to $\tau = 74$ ns [cf. Fig.3(f)], the shortest achievable pulse delay due for the experimental setup used in this study. This indicates that $T_2 \sim T_2^*$ and, thus, that $T_2$ of charge carriers is much shorter in Alq$_3$ compared to previously studied π-conjugated polymer layers [8,34,36].

In Figure 4(a) the results of continuous wave multi-frequency EDMR spectra [8,27,28,34,38], measured over a frequency range from 100 MHz to ~17 GHz, are displayed. The figure shows the zeroth derivative spectra over the whole frequency range as directly obtained from amplitude-modulated lock-in measurements, as well as through numerically integrated and baseline-corrected raw data sets obtained from magnetic field-modulated EDMR spectra for excitation frequencies above 5.55 GHz. The spectra have been normalized to the same maximum amplitude, vertically offset, and horizontally shifted such that the resonance maxima coincide in order to allow for a direct comparison of line shape and width. We find that the FWHM line width increases with excitation frequency from 1.8 mT at the lowest up to 2.9 mT at the highest frequency. One remarkable feature found in Alq$_3$ is that, while the lower frequency spectra [below ~1 GHz, cf. Fig. 4(b), lower panel] overlap to a large degree, the line widths at higher excitation frequencies differ widely, even for spectra recorded from nominally identical devices at the same excitation frequency under nominally identical conditions. This is illustrated in Fig. 4(b), upper panel, where several cw spectra recorded at 10 GHz are plotted; the differences and the variation in line width is quite pronounced. In each case, the overall EDMR line shape can be described as a superposition of two Gaussian lines with root-mean-square line widths $\sigma_1$ and $\sigma_2$, corresponding to the respective charge carriers [28]. In Fig. 5, we plot the wide [cf. Fig. 5(a)] and narrow [cf. Fig. 5(b)] cw line width measured at 200 MHz (blue data points) and at 10 GHz (red data points) as a function of OLED current, i.e. at different device operating points, adjusted by changing the device bias [cf. Fig. 1(a)], measured on the same OLED device. $\sigma_1$ and $\sigma_2$ as well as the respective error bars are determined from a nonlinear least-squares fit to a double-Gaussian model. We find that the line widths measured at 10 GHz depends only weakly on the device operating point, whereas at 200 MHz, both line widths vary when the device current is increased.

## IV.    DISCUSSION

The general trend of the device current as a function of time and magnetic field following a pulsed MW excitation, shown in Fig. 2(a), is qualitatively similar to that seen in other OLED materials [8,16,29,32]. Harneit *et al.* [32] have modeled the observed field-dependent current transients using the product of a field-dependent term $Y(B_0)$ and a time-dependent term $I(t)$. The field dependent term

$$Y(B_0) = \frac{1}{\sqrt{2\pi}\Delta B_{1/2,1}} e^{-\left(\frac{B_0 - B_{c,1}}{\sqrt{2}B_{1/2,1}}\right)^2} + \frac{\left(\frac{1}{r}\right)}{\sqrt{2\pi}\Delta B_{1/2,2}} e^{-\left(\frac{B_0 - B_{c,2}}{\sqrt{2}B_{1/2,2}}\right)^2} \tag{1}$$

reflects the double-Gaussian spectral line shape, where $B_0$ is the magnetic field, $B_{c,1}$ and $B_{c,2}$ are the line centers, $\Delta B_{1/2,1}$ and $\Delta B_{1/2,2}$ the FWHM line widths of both Gaussians and $r$ being their weight ratio [27]. The time-dependent term

$$I(t) = \left(1 - e^{-\frac{(t-t_d)}{t_s}}\right) \sum_{j=1}^{2} I_j e^{-\frac{(t-t_d)}{\tau_j}}, \tag{2}$$

is described by the parameters $t_s$, the rise time of the detector, $t_d$ is the pulse trigger delay, as well as $I_j$ and $\tau_j$, which are the multi-exponential weights and time-constants. These parameters govern the dynamics of the spin-dependent processes that is described by these term, i.e. weakly spin-coupled electron-hole pairs that are generated under bipolar charge-carrier injection conditions, which undergo spin-dependent recombination transitions. These so-called polaron pairs have been found to dominate EDMR (and ODMR) response in organic materials—mostly conjugated polymers—with weak spin-orbit coupling and strong spin-selection rules [9].

We attempt to describe the field-dependent current transients shown in Fig. 2(a) with such a model using a non-linear least-squares fitting procedure on the entire dataset in order to extract the various parameters. While describing the data to some degree, the model does not allow for a residual-free fit of the experimental dataset, as shown in Fig. 2(d) which plots the difference between the best fit result and the measured data. Fig. 2(b) shows a magnetic field slice of the two-dimensional residual dataset, corresponding to a time $t = 25.10$ μs, as indicated by a vertical dashed line in Fig. 2(a,c). This plot reveals features that are inconsistent with a simple polaron pair process, as the model would require the field-dependent slice to vanish between the signal enhancement and the signal quenching at a time t ≈ 25.10 μs due to the product of the field-dependent term with a vanishing time-dependent term. As this residual is not vanishing, we conclude that, while the observed spin-dependent currents are responsible are largely consistent with a polaron-pair model, there are also smaller but significant deviations from this model, indicating the presence of one or more spin-dependent processes that contribute to the recombination current in Alq$_3$ [13].

Figure 3(a,c) displays a least-squares fit of the observed Rabi oscillations to a model function

$$A \sin(\omega_1 t + \Phi) e^{-\frac{t}{T_2^*}} + a + bt + ct^2 + dt^3 \tag{3}$$

which describes the Rabi oscillations as well as the decay of the envelope in time domain, also taking a polynomial baseline $a + bt + ct^2 + dt^3$ into account. From this procedure, we obtain the dephasing time $T_2^*$ as well as the oscillation frequency $\omega_1$, which reflects the power of the MW excitation and $\Phi$ is the phase. We find $T_2^*$ to be 20 ns ± 0.8 ns at room temperature [cf. Fig. 3(a)] and 55 ns ± 15 ns at 5 K. If the rapid dephasing—in particular at room temperature—is not due to spin-dephasing but rather spin-relaxation, it could be attributed to the increased SOC in Alq$_3$ that is caused by the presence of Al. In Figure 3(b,d) we show the frequency spectrum of the datasets in Fig. 3(a,c), obtained via numerical Fourier transformation. In the frequency spectra, both at room temperature and at 5 K, only the fundamental frequency component at the spin-1/2 Rabi frequency is discernible, and higher frequency components are not visible. This is most likely due to the rapid dephasing of the spin oscillations, which limits the width of the fundamental peak and obscures potential additional frequency components. In particular, a spin-beating signal at twice the fundamental Rabi frequency due to the simultaneous excitation of both carrier spins under strong MW drive [11] is not visible at room temperature and at 5 K, which could be attributed to the rapid dephasing and, therefore, the resulting limited frequency resolution of Rabi-frequency spectrum. This however, would also imply that additional frequency components, which may be due to additional spin-dependent recombination processes in another spin manifold [7], may be present but are obscured by the same effect, i.e. the large width of the fundamental spin-1/2 Rabi peak. We therefore, conclude, as for the data sets shown in Fig. 1, that spin-dependent charge carrier recombination in Alq$_3$ appears to be dominated by a polaron-pair process, yet the presence of additional, qualitatively different spin-dependent processes cannot be excluded.

In order to corroborate the hypothesis expressed in the last paragraph, we consider the results of electrically detected Hahn-echo experiment displayed in Figure 3(f) which shows unresolved echo shapes, even at the shorted available pulse delay of $\tau = 74$ ns. The data shows that $T_2$—which describes the decay of echo amplitude with $\tau$—must be very short, which support the hypothesis that increased SOC causes shorter spin-spin relaxation times. Without a discernible echo signal, we cannot establish a number for $T_2$, however, together with the data obtained for $T_2^*$, upper and lower boundaries can be established. Assuming an echo signal-to-noise ratio of less than 1—which is the threshold at which the echo signal becomes discernible—together with the maximum oscillation amplitude in Fig. 3(a) we estimate 20 ns < $T_2$ < 70 ns at room temperature.

The continuous wave multi-frequency analysis using the EASYSPIN toolbox [40] shown in Figure 4(a) exhibits the general trend that has been observed in other OLED materials as well [8,27,28,34,38]: the overall line width remains constant at the lowest excitation frequencies, and exhibits a substantial broadening as the excitation frequency increases. This characteristic is due to the interplay between frequency-independent and frequency-dependent inhomogeneous line width contributions [27]. The frequency-independent width contributions arise from the unresolved hyperfine coupling between the charge-carrier spins and the nuclear spins of hydrogen—which is abundant in organic materials—and can be modeled as a slow-varying, i.e. static distribution of local random magnetic fields, as the nuclear spin polarization is assumed to be negligible within the magnetic field range applied in this study. These resulting *hyperfine* fields lead to an inhomogeneous shift of the individual charge carrier magnetic resonance fields and thus, to a random line broadening. This broadening is different for each of the two charge-carrier species due to the different microscopic localization [8] of the respective molecular orbitals within the Alq$_3$ molecule. At high magnetic field and, thus, excitation frequencies, the influence of the distributions of the charge carriers' *g*-factors, also referred to as *g*-strain, become increasingly relevant for the charge carrier Larmor frequencies and, thus, for the distribution of the Larmor frequency differences within the charge carrier pairs (the so-called *Δ*g-effect) [38]. Also, the anisotropies of the electron and hole *g*-factors, described by *g*-tensors, affect the resonance lines with increasing magnetic field [28]. All these effects lead to an additional inhomogeneous line broadening that scales linearly with excitation frequency [27,33,38]. In organic materials, this broadening mechanism is found to be generally weak due to the abundance of mostly light atoms (hydrogen, carbon, oxygen) and, thus, small SOC, which causes only weak effective *g*-factors. However, at elevated excitation frequencies, these contributions will eventually outweighs the frequency-independent static hyperfine fields.

In Alq$_3$ we find a very uniform line width at excitation frequencies below 1 GHz, see Fig. 4(b), lower panel, for nominally identical devices measured under similar conditions. This suggests that the line width contribution due to hyperfine fields is approximately constant and independent of the microscopic details of the sample morphology. On the other hand, the SOC-dominated lines vary considerably, even in the case of measurements under nominally identical conditions. This can be seen in Fig. 4(b), upper panel, where several spectra measured at 10 GHz are plotted. The spectra exhibit a fairly wide range of different line widths, which suggests that the effect of SOC on the g-factor depends strongly on the nature of the individual devices. In addition, changes of the OLED operating point appear to influence the line width at low excitation frequencies, i.e. at 200 MHz, whereas at higher frequencies, i.e. at X-band, the line widths are rather constant (cf. Fig. 5), indicating that changes of the quasi Fermi level, enable different electronic states (e.g. different molecular orbitals) to contribute to the observed spin-dependent recombination current. A change of the device current leads to a significant modification of the unresolved hyperfine coupling. This effect is much more pronounced at lower excitation frequencies (blue data points in Fig. 5), where the overall line width is dominated by these hyperfine fields, and much less pronounced at higher frequencies (red data points in Fig. 5), where SOC dominates confirming previous observations reported in Ref. [22], where changes in the line widths of Alq$_3$ OLED devices as a function of current density $J$ are reported and are interpreted to have different effect with change of the buffer layer between the cathode/polymer interface. Given the additional evidence found here that this effect is most pronounced in the low-magnetic field domain, yet less pronounced in the high magnetic field domain, we can now attribute this effect do a variation of the hyperfine field strength under different bias conditions, rather than a change of SOC.

## V. CONCLUSIONS

The study of spin-dependent electronic transitions of Alq$_3$ under bipolar injection conditions, i.e. in Alq$_3$ based OLEDs under forward bias conditions, reveal the presence of more than one spin-dependent recombination process. From the analysis of the dynamic response of the device current following a magnetic resonant pulse under constant device bias, we determine that spin-dependent recombination in these devices is dominated by a process involving intermediate electron-hole pairs—similar to previously observed polaron pair processes in OLEDs based on π-conjugated polymers. The significant residual of the fit the dynamics of the observed spin-resonantly controlled spin-dependent recombination currents with the predictions for a single polaron pair process points towards the existence of an additional, independent spin-dependent process whose particular nature has not yet been corroborated. The inability to observe Hahn-echoes with pulsed EDMR as well as the observed rapidly dephasing spin Rabi-beat oscillations are attributed to a short coherence time $T_2 <$ 70 ns and an even shorter dephasing time $T_2^*$ (~10 ns), the latter both at room and low (5 K) temperatures, indicating comparatively strong spin-orbit coupling as the cause. Multi-frequency continuous wave EDMR spectra reveal that charge carrier line widths are governed by local hyperfine field distributions for low frequencies (<500 MHz) and a surprisingly large variation in the SOC-dominated line widths at higher excitation frequencies, for nominally identical devices under nominally identical experimental conditions. This indicates a much higher degree of morphological disorder in the $Alq_3$ films—compared to conjugated polymer systems—leading to substantial fluctuation in the SOC-induced g-factor distributions. Finally, EDMR spectroscopy under varying bias conditions also reveals a variation of EDMR line width, yet only under magnetic field strengths where hyperfine fields dominate the line shape. Thus, it appears that the device's bias conditions allow a tuning of the hyperfine fields that affect charge carrier recombination.

## ACKNOWLEDGMENTS

This work has been supported by the National Science Foundation, NSF-DMR #1701427.

**FIGURES**

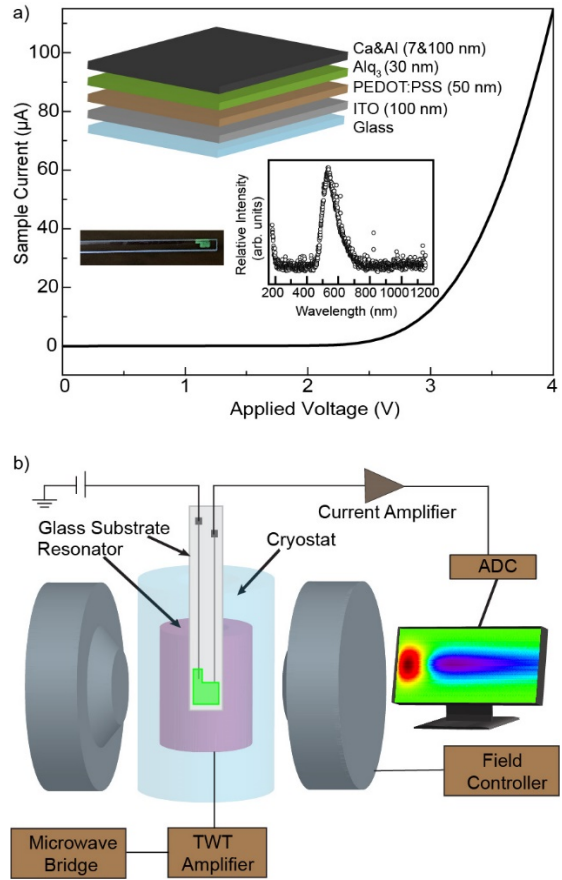

FIG. 1. Alq$_3$ device characterization and experimental setup. (a) Representative I-V curve of one of the studied devices. The insets show a photograph of a device under operation, its luminescence spectrum, and its layer structure on a narrow glass substrate. This structure consists of PEDOT:PSS on ITO as hole injection electrode, Alq$_3$ as the active medium, and Al/Ca as electron injection electrode. (b) Schematic of the pulsed EDMR experimental setup used in this study.

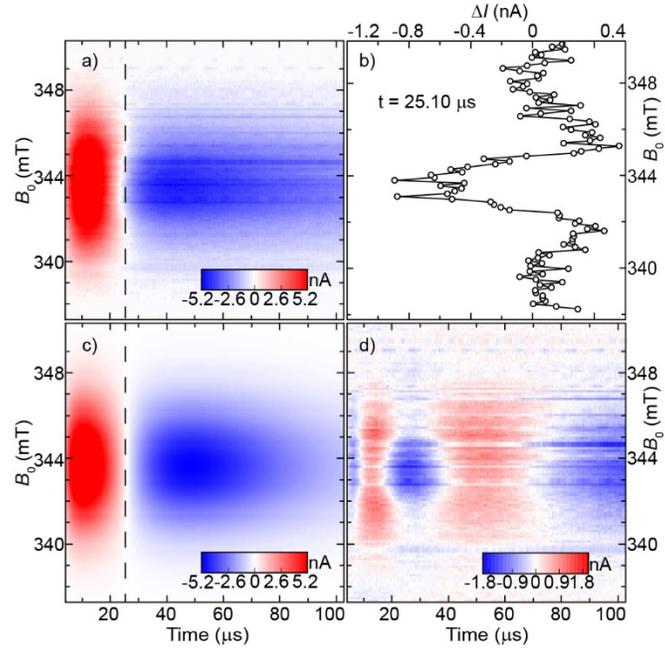

FIG. 2. (a) Plot of the OLED current change following a short microwave pulse (at time t = 0; 400 ns duration, frequency 9.632 GHz, power 1 kW) as a function of time and applied static magnetic field $B_0$. (b) Plot of data from panel (a) recorded at t = 25.10 μs (a time where the current change reverses sign). (c) Least-square fit result of the data in panel (a) using a two-dimensional model taking the spectral line shape and the time dependence into account, as described in Ref. [5]. (d) Plot of the fit residuals, i.e. differences between the measured data in (a) and the fit result in (c). The plot reveals signals outside the fit model.

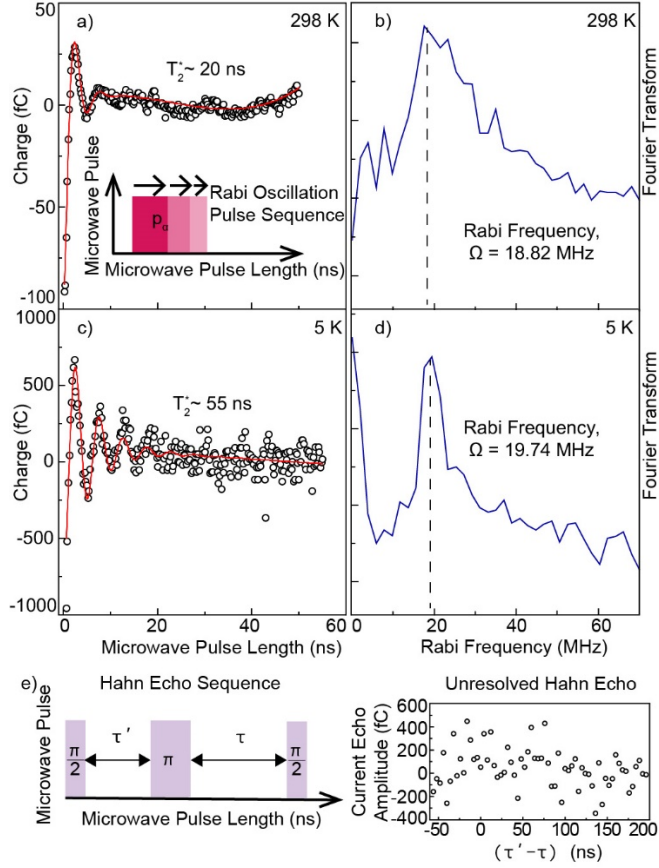

FIG. 3. (a) Measured OLED current (open circles) as a function of pulse duration $p_\alpha$ digitally integrated over an interval of 16.38 μs beginning 2 μs after the pulse at resonance maximum at room temperature. The pulse sequence is illustrated in the inset. The solid line represents a least-squares fit of the experimental data. (b) Fourier transformation of the data in panel (a). (c) Integrated OLED current as a function of pulse duration at resonance maximum at 5 K. (d) Fourier transformation of the data in panel (c). (e) Electrically detected Hahn echo measurements at room temperature. The pulse sequence is shown on the left side, whereas on the right side the echo signal, i.e. the integrated OLED current following the readout-pulse is shown as a function of $\tau - \tau'$ for $\tau = 74$ ns. No echo signal is discernible within the signal-to-noise ratio.

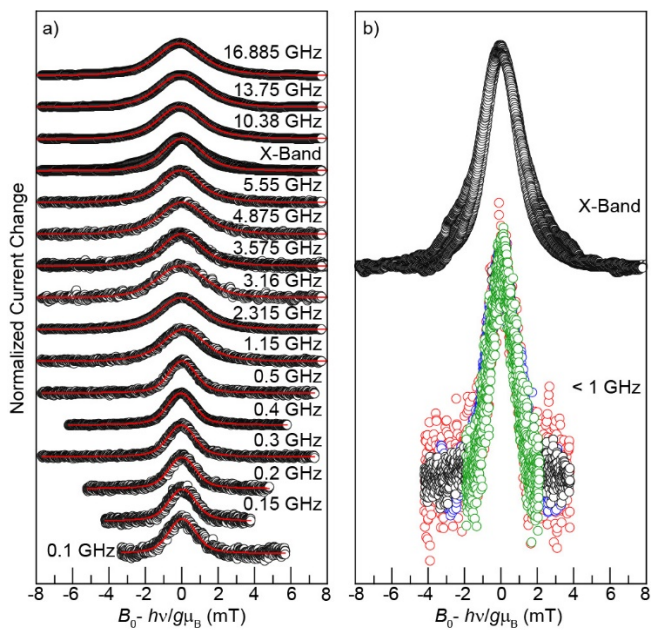

FIG. 4. (a) Multi-frequency continuous wave EDMR spectra measured at excitation frequencies ranging from 100 MHz to 17 GHz. The abscissa is normalized in a way that the resonance maximum occurs at zero. The spectra are normalized to exhibit a comparable amplitude. The solid line corresponds to a least-squares fit to a double-Gaussian model. (b) Comparison of spectra from panel (a) measured at X-band and at excitation frequencies below 1 GHz, respectively. A pronounced broadening with higher frequency/magnetic field is recognizable when X-band data is compared to data below 1GHz. However, for the given noise levels, there are no discernable differences between the line shapes obtained for different frequencies below 1GHz.

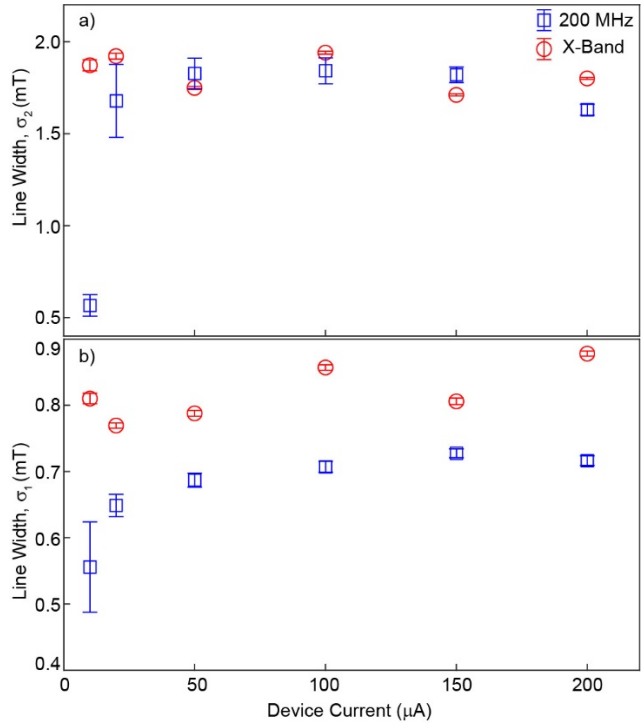

FIG. 5. Root-mean-square line widths for the wide (a) and narrow (b) Gaussian resonance lines measured at 200 MHz and X-Band as a function of OLED current, established from a least-squares fit of a double-Gaussian line shape to the measured spectra. The error bars correspond to the parameter error estimate of the least-squares fitting procedure.